\documentclass[12pt,singlecolumn,twoside]{pnas-new} 

\templatetype{pnasmathematics} 

\usepackage{multibib}
\usepackage{amsmath}

\title{Stability of Hydrides in Sub-Neptune Exoplanets with Thick Hydrogen-Rich Atmospheres}

\author[1,*]{Taehyun Kim}
\author[1]{Xuehui Wei}
\author[2]{Stella Chariton}%
\author[2]{Vitali B. Prakapenka}%
\author[2]{Young-Jay Ryu}%
\author[3]{Shize Yang}%
\author[1,*]{Sang-Heon Shim}

\affil[1]{School of Earth and Space Exploration, Arizona State University, Tempe, AZ 85287, USA.}
\affil[2]{Center for Advanced Radiation Sources, University of Chicago, Argonne, IL 60439, USA.}
\affil[3]{Eyring Materials Center, Arizona State University, Tempe, AZ 85287, USA.}
\affil[*]{Corresponding authors}

\leadauthor{Kim} 

\significancestatement{
Many sub-Neptune exoplanets are believed to have thick atmospheres dominated by hydrogen, surrounding the interiors with mainly silicates and oxides, assuming no reaction between them. 
From high-pressure experiments on molten magnesium oxide in a dense hydrogen liquid, we found that the reaction between hydrogen and the oxide melt results in the formation of hydride and water under the pressure-temperature conditions expected for the atmosphere-interior boundary of sub-Neptunes.
At higher pressures, we found enhanced chemical mixing between magnesium and hydrogen.
The experiments suggest that sub-Neptune exoplanets likely have a mineralogy different from that of rocky planets, and the chemical reaction between the atmosphere and interior can impact the size of sub-Neptunes and the compositions of their atmospheres.
}

\authorcontributions{
T.K. performed the experiments, analyzed data, and wrote the manuscript. 
S.-H.S conceptualised and supervised the project, acquired funding, designed and performed experiments, analyzed data, and wrote the manuscript. 
S.C., Y.-J.R., S.Y., and V.B.P. provided resources for the sample analysis and supervised the experiments.
All authors reviewed the manuscript.
}
\authordeclaration{The authors declare no competing interests.}
\equalauthors{
}
\correspondingauthor{\textsuperscript{*}To whom correspondence should be addressed. E-mail: tkim95@asu.edu and sshim5@asu.edu}


\begin{abstract}
Many sub-Neptune exoplanets have been believed to be composed of a thick hydrogen-dominated atmosphere and a high-temperature heavier-element-dominant core.
From an assumption that there is no chemical reaction between hydrogen and silicates/metals at the atmosphere-interior boundary, the cores of sub-Neptunes have been modeled with molten silicates and metals (magma) in previous studies.
In large sub-Neptunes, pressure at the atmosphere-magma boundary can reach tens of gigapascals where hydrogen is a dense liquid.
A recent experiment showed that hydrogen can induce the reduction of Fe$^{2+}$ in (Mg,Fe)O to Fe$^0$ metal at the pressure-temperature conditions relevant to the atmosphere-interior boundary.
However, it is unclear if Mg, one of the abundant heavy elements in the planetary interiors, remains oxidized or can be reduced by H\@.
Our experiments in the laser-heated diamond-anvil cell found that heating of MgO + Fe to 3500--4900~K (close to or above their melting temperatures) in a H medium leads to the formation of Mg$_2$FeH$_6$ and H$_2$O at 8--13~GPa.
At 26--29~GPa, the behavior of the system changes, and Mg--H in an H fluid and H$_2$O were detected with separate FeH$_x$.
The observations indicate the dissociation of the Mg--O bond by H and subsequent production of hydride and water. 
Therefore, the atmosphere-magma interaction can lead to a fundamentally different mineralogy for sub-Neptune exoplanets compared with rocky planets.
The change in the chemical reaction at the higher pressures can also affect the size demographics (i.e., "radius cliff") and the atmosphere chemistry of sub-Neptune exoplanets.
\end{abstract}

\dates{This manuscript was compiled on \today}
\doi{\url{www.pnas.org/cgi/doi/10.1073/pnas.2309786120}}

\begin{document}

\maketitle
\thispagestyle{firststyle}
\ifthenelse{\boolean{shortarticle}}{\ifthenelse{\boolean{singlecolumn}}{\abscontentformatted}{\abscontent}}{}

\dropcap{T}hick hydrogen-rich atmosphere is likely common in early rocky planets and among large low-density planets. 
In the early stage of planet formation, the accretion of planetesimals forms a core comprised of Mg-rich silicates/oxides and Fe-rich metal alloys \citep{lambrechts2012rapid}. 
As the core becomes sufficiently large  ($>$0.5--0.75~$M(E)$, Earth mass unit), it can accrete the nebular gas (mostly hydrogen), and therefore, planets can develop hydrogen-rich primary atmospheres \citep{lammer2018origin,venturini2020Setting}. 
Earth and Venus in our solar system and the larger versions of them, such as super-Earth exoplanets ($\leq$1.7--1.8~$R(E)$, Earth radius unit), have sufficient core masses to retain a hydrogen-rich atmosphere at least early in their geological history \citep{owen2020Hydrogen}. In such a setting, chemical processes at the boundary between the atmosphere and magma ocean can impact the properties and dynamics of the planets.

The exoplanet surveys for the past 2--3 decades have accumulated mass-radius data for a large number of exoplanets. 
Such data have revealed previously unknown features in the demographics of the exoplanets, providing a new opportunity to advance our knowledge on their origin and evolution. 
One of the significant discoveries from the surveys so far is that planets with sizes between Earth and Neptune are common in our galaxy. 
Among these planets, sub-Neptunes at 1.8--3~$R(E)$ have lower densities relative to rocky planets, suggesting a significant mass of atmosphere for sub-Neptunes.
Sub-Neptunes likely have sufficient core masses to retain hydrogen-rich atmospheres \citep{owen2020Hydrogen}. 
Therefore, models involving the rocky/metallic core with the hydrogen-rich atmosphere have been widely accepted for sub-Neptunes \citep{bean2021Nature}. 
In those planets, particularly larger ones, the atmosphere is likely thick enough that the pressure reaches tens of GPa at the boundary between the atmosphere and the interior.
A thick atmosphere acts as a thermal blanket for the core to maintain a molten state (i.e., magma) for billions of years \citep{vazan2018Contribution, kite2020Atmosphere,zeng2021New}. 
Therefore, the physical and chemical interaction between dense hydrogen and molten silicates/oxides at high pressure-temperature ($P$-$T$) is the key to understand the interiors and atmospheres of sub-Neptunes. 

Until recently, however, most models for these planets assumed that dense hydrogen and magma do not react with each other. 
Some recent models have considered the ingassing of hydrogen in magma in the form of H$_2$ \citep{olson2018Hydrogen} and redox reactions involving Fe and Si \citep{schlichting2022Chemical}  based on low pressure data.  
Recent models for sub-Neptunes have shown that the reactions between the hydrogen-rich atmosphere and magma can produce water and alter the composition of the atmosphere and interior \citep{schlichting2022Chemical,horn2023Reaction}. 
For large low-density planets, the endogenic production in the interior can be an important source for water \citep{horn2023Reaction}. It was proposed that such endogenic production may be important for Earth's water as well \citep{young2023Earth}. It is also possible that potential hydrogen-magma interaction may impact the exoplanet demographics.
For example, the occurrence of sub-Neptunes show a precipitous drop at $3R(E)$, i.e., known as the radius cliff \citep{fulton2018California}.
It was proposed that high pressure expected under the thick atmosphere of a ${\sim}3R(E)$ sub-Neptune may promote efficient physical ingassing of hydrogen to the magma, preventing further growth of the planet beyond the radius cliff \citep{kite2019Superabundance}.

Horn et al.\ \citep{horn2023Reaction} showed experimentally that hydrogen can reduce Fe$^{2+}$ in (Mg,Fe)O to Fe metal when the material is partially molten up to 4000~K at 20--40~GPa. The reduced Fe metal subsequently reacts with H and forms Fe-H alloy. 
The reaction releases O atoms to an H medium, which then react with H to form H$_2$O. 
Therefore, the reaction allows for H ingassing as an alloy component in Fe metal (H$^0$) and hydroxyl ((OH)$^-$ or H$_2$O) dissolved in magma. 
Shinozaki et al.\ \citep{shinozaki2014Formation} showed the dissolution of SiO$_2$ in an H$_2$ liquid at 2--3~GPa and 1500--1700~K, forming Si--H. 
The reaction also produces H$_2$O. 
Despite these important developments, the stability of MgO, one of the dominant components of planetary silicates and oxides, in an H-rich environment is still not well known for the range of $P$-$T$ conditions expected for the atmosphere-interior boundary of sub-Neptunes (0.1--30~GPa and $\geq$2000~K) \cite{stokl2016Dynamical,zeng2021New}. 
Therefore, the persistence of MgO after heating documented in the experiments could be because of the kinetics of the solid-liquid reaction instead of the stability of MgO.
In this study, we report experiments on MgO + Fe metal mixtures in a pure hydrogen medium at 3500--4900~K and 8--29~GPa, conditions close to the melting of MgO \citep{kimura2017Melting} (Fig.\,\ref{1_PT}).
We will discuss implications of the experimental results for sub-Neptune exoplanets.

\section*{Methods}

Thin foils of an MgO + Fe metal mixture were loaded in diamond-anvil cells (DACs).
The foils were propped by MgO grains, as spacers, and pure H$_2$ gas was loaded in the DACs (Fig.\,S1 and Text\,1). 
Experiments were conducted at beamline 13-IDD of GeoSoilEnviroCARS (GSECARS) at the Advanced Photon Source (APS). 
A double-sided laser-heating setup was combined with X-ray diffraction for measurements at high $P$-$T$ \cite{prakapenka2008Advanced}. 
Because of the diamond embrittlement by H, we conducted pulsed laser heating \citep{goncharov2010Xray,fu2023Core,piet2023Superstoichiometric} (Text\,2) to reach temperatures close to the melting of MgO\@.
While the pulse heating method enables us to reach temperatures above 2000~K for the H-rich sample, the heating duration is 0.1--0.2 second, which is not sufficient for measuring X-ray diffraction patterns with low noise.
Furthermore, our heating targets melting.
When a crystalline sample undergoes melting, its sharp and intense diffraction lines disappear, and broad diffraction features from melt appear.
In a DAC, such broad diffraction features are difficult to separate from the strong background caused by Compton scattering from diamond anvils.
Furthermore, melting in an H fluid can result in the removal of materials at the heating center (therefore a lack of diffraction at the heating center \citep{fu2022Stable}; also found in our experiments as shown in Fig.\,S2a).
Therefore, we focus on analyzing the diffraction patterns measured before and after laser heating at high pressures in this study.

Thermal radiation spectra were measured for the both sides of the sample for temperature measurements.
Pressures at high temperatures were calculated from the measured MgO volumes combined with its equation of state \citep{dorogokupets2007Equations}. 
Diffraction measurements (a beamsize of $3{\times}2\,\mu$m$^{2}$) were performed at the center of the heated area as well as its adjacent areas.
Diffraction images were measured for 0.1--0.2~second at high temperatures and 30 seconds after heating at 300~K\@. 
Raman spectroscopy was conducted at GSECARS \citep{holtgrewe2019Advanced} and ASU using a 532-nm laser as an excitation source (Text\,3). 
The samples were recovered and prepared in a focused ion beam (FIB) instrument (Helios 5 UX, Thermo Fisher Scientific Inc.) at ASU\@.
The recovered samples were analyzed in scanning transmission electron microscopy (ARM-200F, JEOL co., at ASU) for chemical composition and imaging. 
A detailed description of our methods can be found in SI Appendix.


\section*{Results}

\begin{figure}[hb] 
\includegraphics[width=0.6\textwidth]{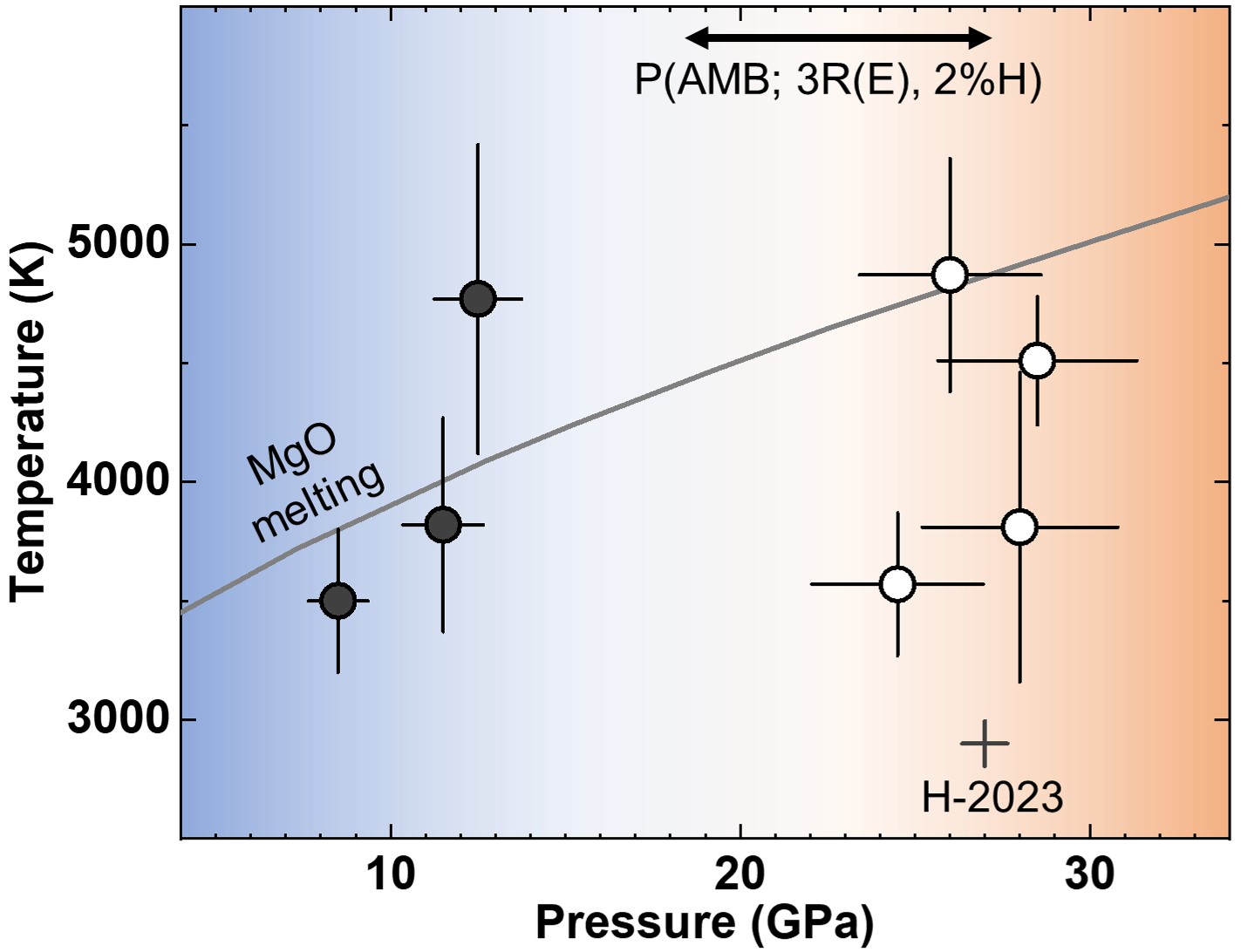}
\caption{Pressure and temperature conditions (circles) of the experimental runs (Table\,S1). 
Black and white circles represent the presence and absence of Mg$_2$FeH$_6$, respectively.
A color gradient in the background indicates a possible change in the chemical behavior of the system at 15--25~GPa. 
The arrow at the top shows the estimated range for the pressure at the atmosphere-magma boundary $P(\mathrm{AMB})$ in sub-Neptunes with masses of $9.2{-}13.6M(E)$ for a radius of $3R(E)$ and 2\% hydrogen (see SI Appendix). 
$R(E)$: Earth radius. $M(E)$: Earth mass.  The MgO melting curve is from \citep{kimura2017Melting}.
H-2023: a data point for MgO + Fe in H from \cite{horn2023Reaction}. 
}\label{1_PT}
\end{figure}

\begin{figure}[hb] 
\includegraphics[width=0.6\textwidth]{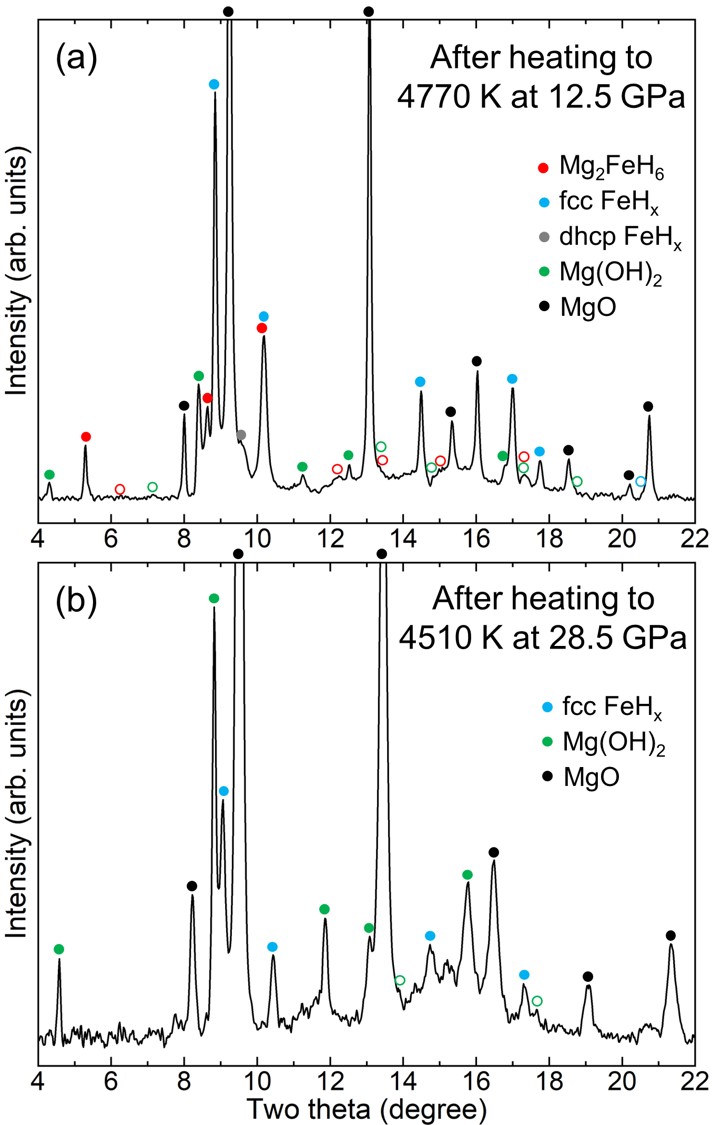}
\caption{X-ray diffraction patterns measured after the melting of MgO + Fe in H$_2$ at (a) 4770~K and 12.5~GPa, and (b) 4510~K and 28.5~GPa. 
The colored circles above the peak indicate the peak positions of the observed phases in the legend. 
If peaks are either weak or not observed, we used open circles. 
The Le Bail analysis of pattern (a) is presented in Fig.\,S6.
The wavelength of the X-ray beam is 0.3344\,\AA.
SEM analysis of the recovered sample for (a) is presented in Fig.\,S2.
}\label{2_XRD}
\end{figure}

At 8--13~GPa after heating, XRD patterns showed the peaks from Mg$_2$FeH$_6$, Mg(OH)$_2$, face-centered cubic (fcc) FeH$_x$, and MgO (Figs\,\ref{2_XRD}a, S3, and S4a). 
As shown in previous high-pressure studies, hydrogen alloys with liquid Fe metal at high $P$-$T$ and form FeH$_x$, $\mathrm{Fe + 0.5}x\mathrm{H_2 \rightarrow FeH}_x$, which is also confirmed here \citep{okuchi1997Hydrogen,sakamaki2009Melting,piet2023Superstoichiometric} (in our experiments, $x = 1.0{-}1.2$; Fig.\,S5). 
Cubic Mg$_2$FeH$_6$ perovskite phase appeared after an MgO + Fe mixture was heated to temperatures above 3500~K, which is close to or above the melting temperatures of MgO (Figs\,\ref{2_XRD}a, S3, and S4a). 
Three intense peaks of Mg$_2$FeH$_6$ were clearly identified in diffraction images (Fig.\,S3). 
In the recovered samples, we often found a hole at the heating center (Fig.\,S2a).
As shown in a previous study \cite{fu2022Stable}, convecting H fluid can remove some melt from the hottest spot and in case the hole becomes sufficiently large, it can be observed in the LHDAC samples with hydrogen.
Therefore, the observation of the holes in some samples can indicate melting.
Horn et al. \citep{horn2023Reaction} studied the same system but only at pressures above 20~GPa.
Shinozaki et al. \citep{shinozaki2013influence} found MgO does not decompose in H$_2$ at 1500~K and 15~GPa.
Therefore, the formation of Mg$_2$FeH$_6$ from MgO + Fe + H$_2$, which implies the dissociation of the Mg--O bond, likely requires heating to sufficiently high temperatures at pressures lower than 15~GPa (Fig.\,\ref{1_PT}): \begin{equation}
\mathrm{2MgO + Fe + 5H_2 \rightarrow Mg_2FeH_6 + 2H_2O}.\label{eq-Mg2FeH6-formation}
\end{equation}

Reaching to the melting temperature of MgO ($\sim$4000--5500~K) in an H medium at high pressure is extremely challenging \citep{horn2023Reaction}. 
In this study, the experimental setup was optimized for achieving temperature as close to the melting of MgO in the presence of H$_2$, although this heating method makes it challenging to measure diffraction patterns at simultaneously high $P$-$T$ (Text 2). 
Previous studies have shown that the breakdown of the Fe--O bond \citep{efimchenko2019Destruction,horn2023Reaction} and the Si--O bond \citep{shinozaki2014Formation} in the presence of hydrogen. 
However, to our knowledge, none of the previous studies have reported the breakdown of the Mg--O bond and the formation of Mg,Fe-hydride at the $P$-$T$ conditions relevant to the atmosphere-magma boundary of sub-Neptunes.

\begin{figure}[hb] 
\includegraphics[width=0.5\textwidth]{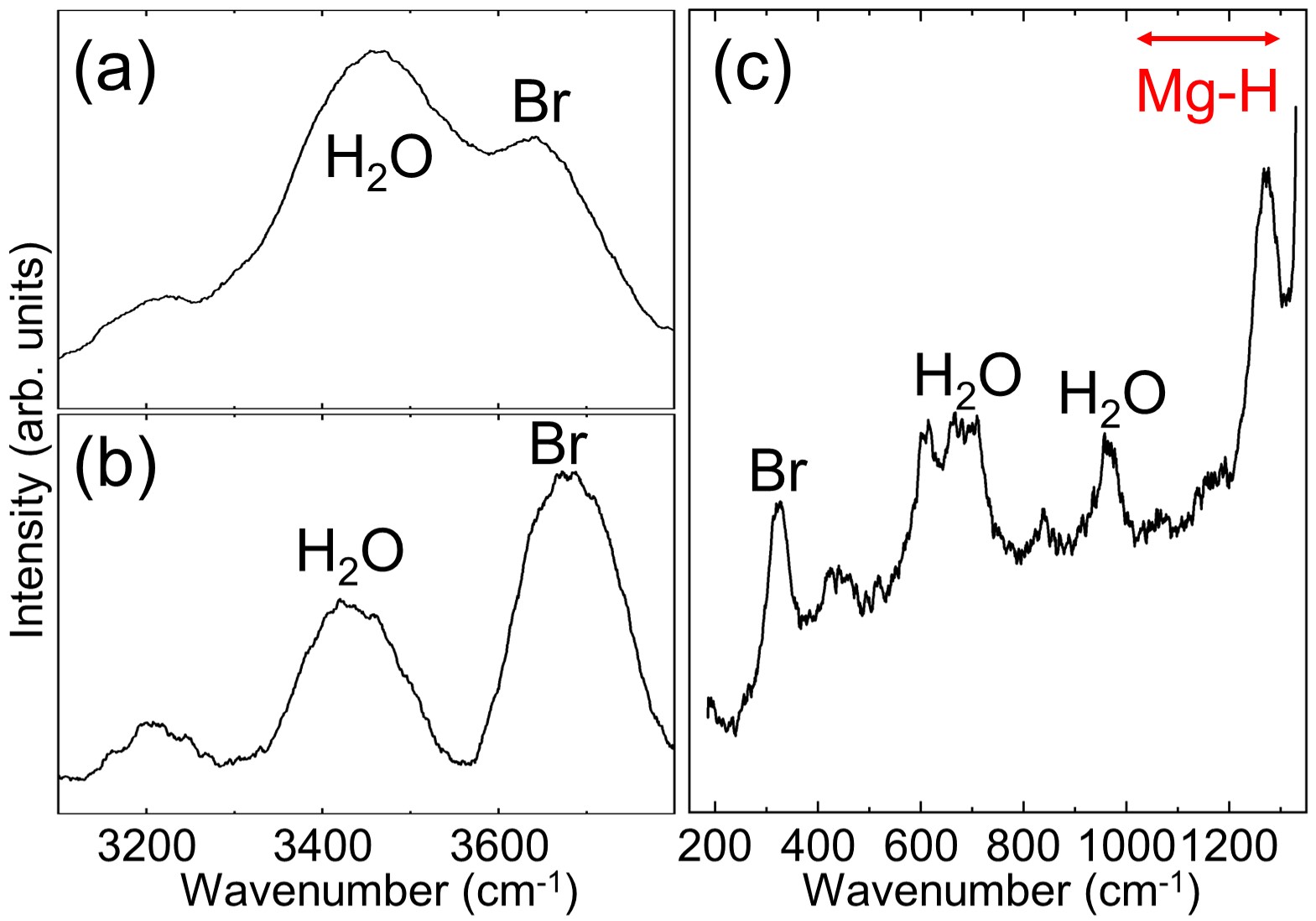}
\caption{Raman spectra measured at (a) 7.0~GPa for the sample heated at 12.5~GPa and 4770~K (\#31-005), and (b) 11.5 GPa and 3820 K (\#31-003), and (c) 22.0~GPa for the sample heated at 26.0~GPa and 4870~K (\#26-009) (Table\,S1). 
SEM data for the sample in (c) is presented in Fig.\,S12.
H$_2$O: H$_2$O-VII and Br: Mg(OH)$_2$.}
\label{3_Raman}
\end{figure}

Because of the Mg--O bond break, some O atoms can be released from MgO and react with either Fe or H in the system.
The XRD patterns do not show any peaks of FeO and Fe$_2$O$_3$ (Figs\,\ref{2_XRD}a, S3, and S4a). 
Under an H-rich environment, iron oxides (FeO and Fe$_2$O$_3$) can be easily reduced to Fe metal by H$_2$ at 2000--3000~K and 25~GPa \citep{horn2023Reaction}, which explains no observation of iron oxides in our experiments.
The diffraction peaks of H$_2$O are difficult to identify because of the peak overlap with fcc~FeH$_x$ expected at this pressure range.
In addition, H$_2$O peaks have much weaker diffraction intensity.
Related to H$_2$O, however, Mg(OH)$_2$ peaks were observed (Figs\,\ref{2_XRD}a, S3, and S4a). 
Mg(OH)$_2$ can form through a reaction between MgO and H$_2$O during temperature quench at lower temperatures: 
MgO + H$_2$O (from Eq.~\ref{eq-Mg2FeH6-formation}) $\rightarrow$ Mg(OH)$_2$.
Therefore, its appearance indicates the formation of H$_2$O as predicted in Eq.\,\ref{eq-Mg2FeH6-formation}. 
In the laser-heated area, the O--H vibrational modes of H$_2$O-ice VII were detected in Raman spectroscopy at high pressure and 300~K after temperature quench (Figs\,\ref{3_Raman}a and \ref{3_Raman}b). 
Raman spectroscopy also detected the O--H vibration from Mg(OH)$_2$.

At 26--29~GPa, XRD patterns showed the formation of fcc FeH$_x$ and Mg(OH)$_2$, whereas Mg$_2$FeH$_6$ peaks were not found (Figs\,\ref{2_XRD}b, S4b, and S4c). 
As discussed above, the appearance of Mg(OH)$_2$ requires H$_2$O formation and therefore the release of O from MgO through the breakage of some Mg--O bond (Eq.\,\ref{eq-Mg2FeH6-formation}).
Similar to the case of Fe$^{2+}$ in (Mg,Fe)O \citep{horn2023Reaction}, it can be hypothesized that O is released by Mg$^{2+}$ reduction to Mg$^0$ metal.
It is also possible that Mg--O bond breaks down, but Mg remains as Mg$^{2+}$.
Diffraction lines of Mg metal or its hydride forms were not observed at high pressures after heating, thus Mg$^{2+}$ was unlikely reduced to Mg metal. 
If Mg$^{2+}$ dissolves in H$_2$ liquid during melting, it would not create any diffraction peaks. 
Related to the system we study here, a recent study showed that MgO can dissolve into H$_2$O fluid at $\sim$30~GPa and above 1500~K \citep{kim2021Atomicscale}. 
Similarly, Mg$^{2+}$ and H may form an Mg--H bond.
In our experiments, when the sample was decompressed to $\sim$1~GPa, several diffraction spots appeared in XRD images. 
Those spots were detected at the same two theta angle at different azimuthal angles (Fig.\,S7). The two-theta angle does not match with any diffraction peaks expected for Mg(OH)$_2$, MgO, Fe metal, and fcc FeH$_x$.
The observed peak is possibly from the crystallization of a component dissolved in H$_2$ liquid at high pressures during decompression. 
However, no more clear diffraction features were observed and therefore we could not identify the phase with XRD alone.
%

After heating at 26--29~GPa, an intense O--H vibration from Mg(OH)$_2$ and H$_2$O was observed at 2800--3700~cm$^{-1}$ as well as the lattice modes of Mg(OH)$_2$ \citep{duffy1995Structure} and H$_2$O \citep{zha2016New} at 200--1300~cm$^{-1}$ (Figs \,\ref{3_Raman}c, S8, and S9). Upon decompression, the peak shifts to lower wavenumbers but remains in the range (Fig.\,S10). 
The Mg--H stretching vibration exists between 900 and 1300~cm$^{-1}$ at 1~bar \citep{shantilalgangrade2017Dehydrogenation} (Figs\,S9 and S10).
At 1~GPa, some additional Raman peaks appeared.
The mode frequencies of these peaks agree well with those expected for the lattice vibrational modes of $\gamma$-MgH$_2$ \citep{kuzovnikov2013Raman} (Figs\,S9 and S10b), although some level of intensity contribution from weak H$_2$O Raman modes cannot be ruled out \cite{zha2016New}. 
From these observations, we interpret that some Mg$^{2+}$ may dissolve into H$_2$ fluid (i.e., Mg--H) at pressures above 25~GPa and $\sim$3500~K:
\begin{equation}
\mathrm{MgO + 2H_2 \rightarrow MgH_2 + H_2O}.\label{eq-Mg-H-formation} 
\end{equation}
The dissolved Mg--H precipitates with decompression because of its lowering solubility in H$_2$ at lower pressures.

In our experiments, a complete conversion of MgO to MgH$_2$ was not observed. 
The total heating duration was 0.1--0.2 second and it could not be extended any further because of the possible mechanical failure of diamond anvils by H\@ (Text\,2).
The heating duration may not be sufficient for the completion of the reaction (Figs\,\ref{2_XRD}, S3, and S4). 
It is also possible that H$_2$O, produced from the reaction, makes the condition more oxidizing than the initial redox condition from a pure H$_2$ medium. 
Therefore, the reaction (Eq. \ref{eq-Mg-H-formation}) may not proceed any further.


\begin{figure}[hb] 
\includegraphics[width=0.9\textwidth]{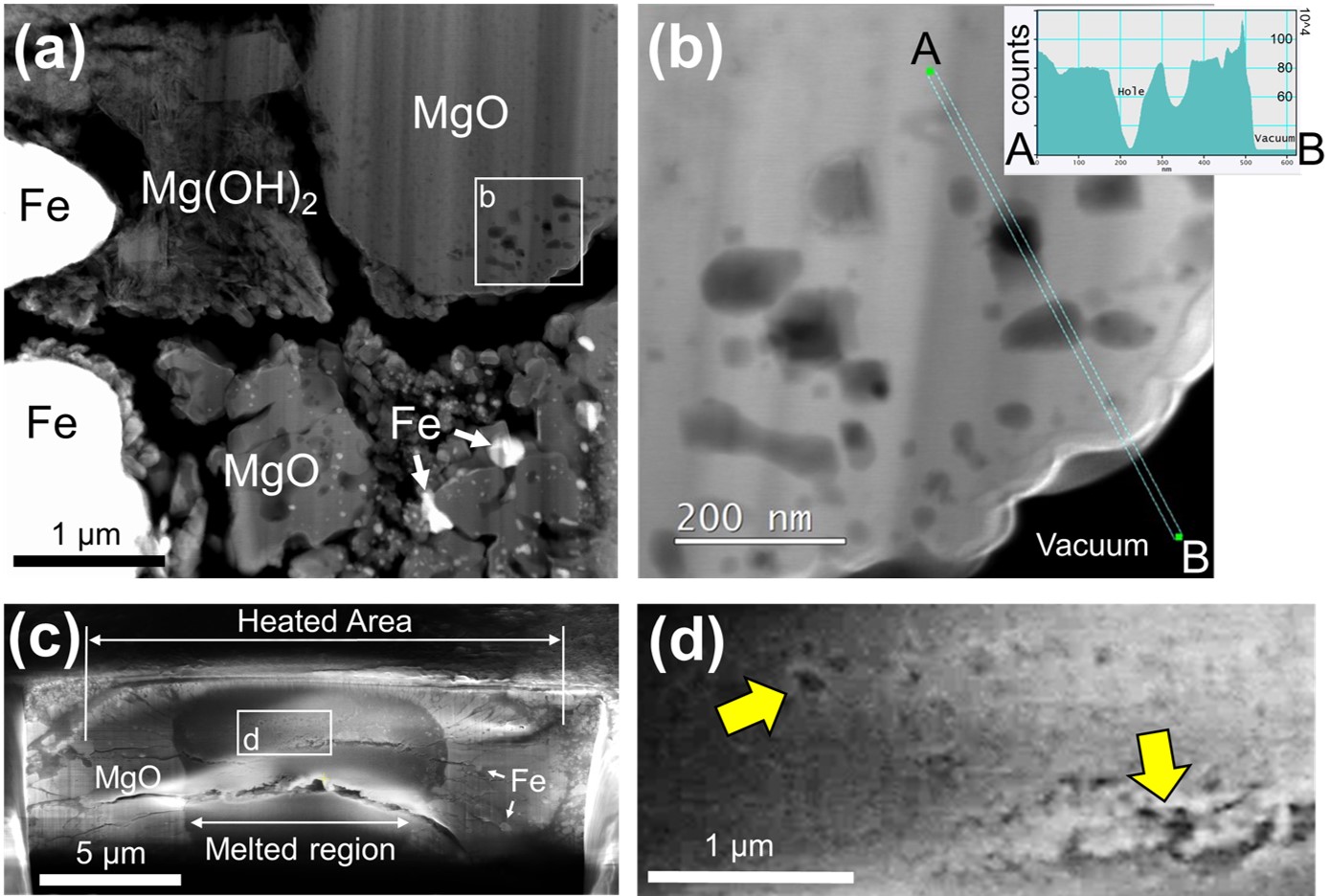}
\caption{STEM (a and b) and SEM (c and d) images of the heated areas in the samples recovered from run~\#31-003 (11.5~GPa and 3820~K) and \#26-009 (26~GPa and 4870~K), respectively (Table\,S1). 
(a) An FIB section shows some of the phases found in XRD\@. 
The small bright grains are bcc Fe metal (labeled as Fe) converted from fcc FeH$_x$ during decompression \citep{badding1991high}.
The diffraction pattern for the converted Fe metal can be found in Fig.\,S7. 
The Fe grains in the left side are not heated. 
(b) Bubble-shaped structures are found along the edge of MgO grans (a white box in (a)).
An intensity line profile between point A and point B is shown in the inset. 
(c) A cross section of the laser-heated area in the sample foil was made in FIB to reveal the structures.
The center of the heated area contains more MgO and the surrounding area contains more Fe metal.
Some Fe metal may migrate during melting along the radial thermal gradients. 
(d) A magnified image (a white box in (c)) reveals the bubble-shaped structures (highlighted by yellow arrows) within the heated area, similar to the ones observed in (b) but in larger sizes (note the scale difference).
}\label{4_TEM}
\end{figure}


The recovered samples were analyzed in STEM (scanning transmission electron microscopy) and SEM (scanning electron microscopy). 
The samples are fractured and mechanically weak because of the O release, melting in a fluid H medium, and hydrogen transition from solid/liquid to gas during decompression (Figs \,\ref{4_TEM}a, \ref{4_TEM}c, S2, S11, and S12). 
Therefore, it was difficult to recover all the samples studied at the synchrotron.
For the sample recovered from 11.5~GPa, bubble-shaped structures were observed near the edge of MgO grains (Fig.\,\ref{4_TEM}b). 
The size of the structure ranges between 50 and 100~nm.
Inside the bubble-shaped area, intensity was found to be as low as that found at the vacuum (Fig.\,\ref{4_TEM}b).
The intensity in an STEM annular dark-field (ADF) image is proportional to $Z^{2-x}$, where $Z$ is the atomic numbers and $x$ is the empirical parameters \citep{yang2019direct}. 
Therefore, the bubble-like structures are either empty or filled with low $Z$ elements, most likely H$_2$ in this experiment.

In a sample recovered from 26~GPa, similar bubble-shaped structures are observed with larger sizes, ranging between 100 and 300~nm (Figs\,\ref{4_TEM}c and \ref{4_TEM}d). It is possible that such bubble-shaped structures were formed from the mechanical capturing of H$_2$ or H$_2$O (formed by the reactions) fluid in the melt during temperature quench.
Alternatively, such structures can form from the separation of two components that were once miscible at high temperature.
In this case, because of the much lower mutual solubility at lower temperatures, the two components can separate into different phases during temperature quenching, resulting in the formation of the bubble-shaped structure.
In the case of the albite--H$_2$O system, they become miscible with each other at 0.73~GPa and 1673~K \citep{paillat1992solubility}.  
During temperature quench, because of the low solubility of H$_2$O in albite at lower temperatures, the exsolved H$_2$O can form bubble-shaped structures in the albite crystal.
Therefore, if the same process occurs for the MgO--H system studied here, H once mixed with MgO melt may form the bubble-shaped structures during temperature quench.
At 25~GPa, the size of the bubble-shaped structure is larger and Mg--H mode is observed in H$_2$, indicating that the mutual solubility during heating is much greater at the higher pressures (Figs\,\ref{4_TEM}d and S12b).

Horn et al. \cite{horn2023Reaction} conducted experiments on (Mg,Fe)O in H at similar $P$-$T$ range.
They reported the reduction of Fe$^{2+}$ and the formation of Mg(OH)$_2$, but did not report the Mg--H mixing.
However, they did not conduct TEM, SEM, and Raman spectroscopy measurements which are critical for detecting the Mg--H mixing in our study.

\section*{Implications}

In recent studies, possible chemical reactions between hydrogen and silicates/metals have been drawn interest for their potential impacts on planetary processes~\cite{horn2023Reaction,young2023Earth,piet2023Superstoichiometric}.
A high-pressure study \citep{horn2023Reaction} showed that H can induce the reduction of Fe$^{2+}$ in (Mg,Fe)O to Fe metal. 
Because of the electron $(e^{-})$ transfer from H$^0$ to Fe$^{2+}$ (the first two reactions in Eq.\,\ref{eq:Fe-reduction-e-transfer} below), O atoms released from (Mg,Fe)O (the first reaction in Eq.\,\ref{eq:Fe-reduction-e-transfer}) react with H$^+$, resulting in the formation of H$_2$O (the third reaction in Eq.\,\ref{eq:Fe-reduction-e-transfer}, which can be obtained by combining first two reactions). 

\protect\begin{equation}\begin{gathered}
    \mathrm{Fe^{2+}O^{2-}} + 2e^- \rightarrow \mathrm{Fe^0 + O^{2-}} \\
    \mathrm{H^0_2} \rightarrow \mathrm{2H^+} + 2e^- \\
    \mathrm{Fe^{2+}O^{2-}~(magma) + H^0_2~(atm)} \rightarrow \mathrm{Fe^0~(metal) + H^{+}_2O^{2-}~(atm~and~magma)}.
\end{gathered}\label{eq:Fe-reduction-e-transfer}\end{equation}
The endogenically produced water can partition into both magma and atmosphere (``atm'' in Eq.\,\ref{eq:Fe-reduction-e-transfer}).

We found that MgO reacts with H at higher temperatures, breaking Mg--O bonds and forming Mg--H bonds (hydride, H$^{-}$), leading to the formation of Mg$_2$FeH$_6$ at pressures below 13~GPa and the possible dissolution of Mg in H fluid (i.e., Mg--H) at pressures above 24~GPa.
Our Raman measurements showed the formation of H$_2$O (and therefore proton, H$^{+}$).
From these observations, the following chemical reactions with electron transfer can be inferred (the fourth reaction can be obtained by combining first three reactions; note that we are talking about a general case here for the Mg-related components which can be applied to both Eqs\,\ref{eq-Mg2FeH6-formation} and \ref{eq-Mg-H-formation}):
\protect\begin{equation}\begin{gathered}
    \mathrm{Mg^{2+}O^{2-}} \rightarrow \mathrm{Mg^{2+}} + \mathrm{O^{2-}}\\
    \mathrm{H^0_2} + 2e^- \rightarrow \mathrm{2H^-} \\
    \mathrm{H^0_2} \rightarrow \mathrm{2H^+} + 2e^- \\
    \mathrm{Mg^{2+}O^{2-}~(magma) + 2H^0_2~(atm)} \rightarrow \mathrm{Mg^{2+}H^-_2 + H^+_2O^{2-} (atm~and~magma)}.\label{eq:Mg-H-formation-e-transfer}
\end{gathered}\end{equation}

The key difference from the Fe case \citep{horn2023Reaction} include: (1) Mg remains oxidized (the first reaction in Eq.\,\ref{eq:Mg-H-formation-e-transfer}), (2) Mg forms bonding directly with H$^{-}$ (hydride) (the second reaction in Eq.\,\ref{eq:Mg-H-formation-e-transfer}), and (3) electron transfer occurs between hydrogen atoms to form protons which react with O to form water (the third reaction in Eq.\,\ref{eq:Mg-H-formation-e-transfer}).
As Mg is one of the dominant heavy elements in the planetary interiors, the chemical process we identified here can add a significant amount of H$_2$O to the total endogenic H$_2$O in addition to the H$_2$O from the Fe$^{2+}$ reduction found in \citep{horn2023Reaction}.
An important difference between Fe and Mg in the reaction with H in a planetary setting is that after the reduction Fe metal would sink toward the center (because of its high density), and therefore be separated from the atmosphere and magma. 
However, Mg, by forming hydride bonding, would likely remain in atmosphere and/or magma, possibly contributing to radial compositional gradients at the shallower depths in the planets together with endogenic H$_2$O.

\begin{figure}[hb] 
\includegraphics[width=\textwidth]{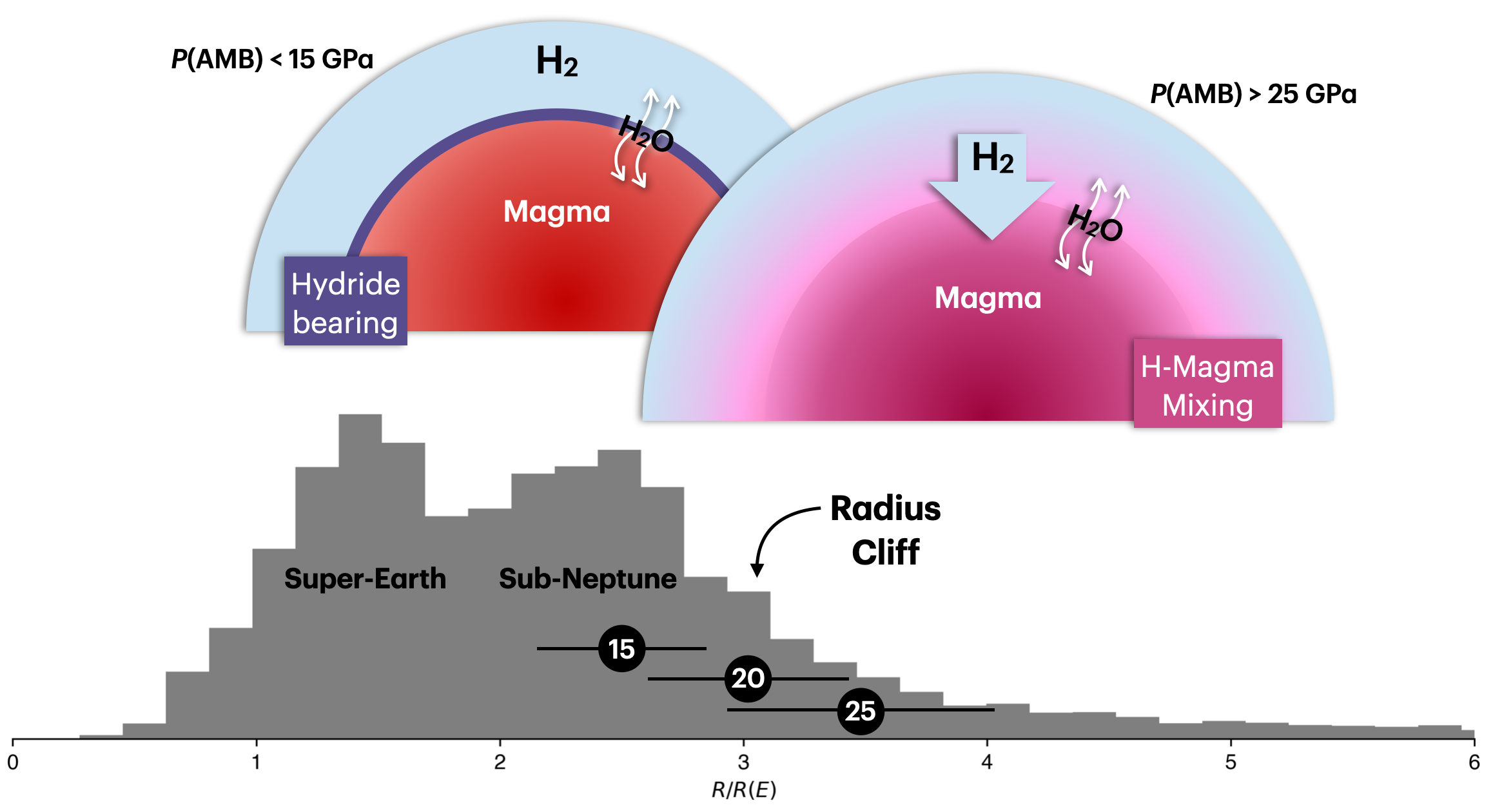}
\caption{An exoplanet size demographics histogram up to $R/R(E) = 6$ ($R$: radius, $R(E)$: Earth's radius; bottom) shown together with the implications of our experiments on the radius cliff (top).
Schematic diagrams for the structures of sub-Neptunes with $P(\mathrm{AMB}) < 15$~GPa and $P(\mathrm{AMB}) > 25$~GPa are shown (AMB: atmosphere-magma boundary). 
In the histogram, three horizontal bars show the estimated radii of sub-Neptunes (2~wt\% H$_2$ atmosphere with an Earth-like interior composition) for different pressures (in GPa) at the atmosphere-magma boundary (see Text\,5 for method).
}\label{5_Density}
\end{figure}

At the $P$-$T$ conditions (8--13~GPa and 3500--4800~K) relevant for the interface between the atmosphere and the magma ocean in H-rich sub-Neptunes, we observed the formation of Mg$_2$FeH$_6$ from the reaction between MgO + Fe and H\@.
For example, for GJ1214b ($6.55M(E)$ and $2.68R(E)$), 8~GPa and 3000~K were estimated for the atmosphere-magma boundary \citep{nettelmann2011Thermal}. 
Mg$_2$FeH$_6$, the material formed under the planetary conditions in our experiments below 13~GPa, has been extensively studied in condensed matter physics and materials science because of its importance for H storage applications \citep{george2009Bulk}.
Our experiments show that Mg$_2$FeH$_6$ can exist in a H-rich planet, i.e., the atmosphere-magma boundary (AMB) of sub-Neptune exoplanets, potentially important for the H storage in a natural setting as well (Fig.\,\ref{5_Density}).

Spectroscopy studies have shown that the valence states of Mg and [FeH$_6$] in Mg$_2$FeH$_6$ are $+2$ and $-4$, respectively \citep{didisheim1984Dimagnesium, retuerto2015Neutron}. 
Therefore, reduction to metal does not occur for Mg under H$_2$, as we found no evidence for Mg metal or alloy in the experiment.
However, our experiments show that by breaking the Mg--O bond MgO + H reaction can still produce H$_2$O (Eqs\,\ref{eq-Mg2FeH6-formation} and \ref{eq:Mg-H-formation-e-transfer}). 
The valence state of Fe in the [FeH$_6$]$^{4-}$ complex has been debated: Fe$^{2+}$ \citep{didisheim1984Dimagnesium} or Fe$^0$ \citep{retuerto2015Neutron}. 
A computational study \citep{zareii2012Structural} showed that the bonding between Fe and H is partially covalent. 
Because of Mg$^{2+}$ in Mg$_2$FeH$_6$ and the higher electronegativity of H than Fe, [FeH$_6$]$^{4-}$ should have some degree of ionic bonding, therefore a negative charge for H\@.
This means hydride bonding should exist regardless of whether the valence state of Fe is 2+ or 0. 
Therefore, our experiments reveal an intriguing possibility of co-ingassing of H$^{-}$ (in Mg$_2$FeH$_6$ hydride) and H$^+$ (in H$_2$O) in a planetary setting.

Our measurements combined with the existing data \citep{jackson1999Elasticity,george2009Bulk} found that Mg$_2$FeH$_6$ has a substantially lower density than MgSiO$_3$ and MgO, which are representative silicate/oxide for the rocky part of planets (Fig.\,S13).
Mg$_2$FeH$_6$ has a perovskite-type structure, similar to the crystal structure of MgSiO$_3$ above 24~GPa (orthorhombic perovskite structure; bridgmanite). 
However, Mg$_2$FeH$_6$ has a vacancy-ordered double perovskite structure \citep{rahim2020Geometric}. 
Bridgmanite, with a normal perovskite structure, has no vacant site between the SiO$_6$ octahedra which are connected with each other by the corner sharing. 
However, in the crystal structure of Mg$_2$FeH$_6$ (space group: $Fm\bar{3}m$, \#225), the FeH$_6$ octahedra, positioned at the lattice points of face-centered cubic, are isolated with each other because of the ordered vacancies \citep{retuerto2015Neutron}. 
Therefore, the crystal structure of Mg$_2$FeH$_6$ and its high H content can explain its low density. 
Given as much as 15--30\% lower density than MgSiO$_3$ between 1~bar and 20~GPa (we compare to MgSiO$_3$ enstatite, a phase stable at the pressure range), Mg$_2$FeH$_6$ likely remains less dense than silicate even at the high temperatures of planetary interiors. 
Therefore, Mg$_2$FeH$_6$ can be buoyant when it crystallizes in the magma ocean or in the solidified silicate rocky layer. 
Such buoyant Mg$_2$FeH$_6$ at the interface between the atmosphere and the interior can be an important agent for maintaining very reducing conditions for the underlying interior. 
It is also feasible that other hydrides may form at the interface \citep{shinozaki2014Formation}. 
If dense hydrides can be formed, and ingassing of hydride (H$^-$) could be extended to much greater depths combined with the convective flow of the interior.
In \citep{horn2023Reaction}, from the higher melting temperature of MgO, it was postulated that MgO may crystallize first from the magma ocean beneath a thick H-rich atmosphere and form a crust.
They further hypothesized that the MgO-rich crust could regulate the interaction between H-rich atmosphere and the magma ocean.
According to our experiments, Mg,Fe hydrides may form at the interface and therefore such an MgO-rich crust unlikely forms.

Mg$_2$FeH$_6$ does not form at pressures above 25~GPa. 
Instead, the Raman data, together with the electron microscopy data, indicate that Mg may dissolve into a H fluid at this pressure range. 
In other words, MgO (possibly in a liquid form) may become (at least partially) miscible with a H$_2$ liquid.

According to the \textit{Kepler} data, planet occurrence drops sharply at ${\sim}3R(E)$ dividing sub-Neptune exoplanets and Neptune-size exoplanets (i.e., radius cliff; Fig.\,\ref{5_Density}) \citep{fulton2018California,hsu2019Occurrence,kite2019Superabundance}.
Kite et al.\ \citep{kite2019Superabundance} proposed that a sudden increase in the H solubility in magma by pressure effects may ingas much more H at the radius cliff, rather than leave much hydrogen for building the atmosphere.
Because of the significant ingassing of H and much smaller molar volume of H expected for the ingassed states compared with pure H$_2$, the atmosphere does not increase as much and therefore the radius of a planet does not increase as much near $3R(E)$, leading to a low population of planets with radius greater than $3R(E)$.

For a sub-Neptune with 2~wt\% H$_2$ atmosphere and an Earth-like interior composition, the pressure at the atmosphere-magma boundary ($P(\mathrm{AMB})$) can be estimated for a planet radius using the equations in \citep{zeng2021New,otegi2020revisited} (Text\,5 for detail).
From the calculation, we found that sub-Neptunes with radii of $2.52$, $3.03$, and $3.48R(E)$ can have pressures of 15, 20, and 25~GPa at the AMB, respectively.
We observed the formation of Mg$_2$FeH$_6$ below 15~GPa.
Therefore, for a smaller sub-Neptune $(R_p \lesssim 2.5R(E))$, Mg$_2$FeH$_6$ may crystallize at the AMB and remain there, because of its lower density than silicate melt (Fig.\,S13).
Its high concentration at the AMB could regulate the H ingassing to the magma ocean beneath.
At pressures higher than 25~GPa, our experiments found that Mg$_2$FeH$_6$ is no longer stable.
Instead, the detection of the Mg--H vibration and the bubble-shaped structures found in the recovered samples point toward much enhanced chemical mixing between Mg and H  at the higher pressure range (Fig.\,\ref{5_Density}b).
Therefore, if a sub-Neptune type planet can grow large enough to $R_p \gtrsim 3R(E)$, the pressure at the AMB can become sufficiently high for the effective chemical mixing between Mg and H, and therefore efficient H ingassing.
If such a process indeed occurs, similar to the proposed process in \citep{kite2019Superabundance}, large sub-Neptune type planet may not grow efficiently beyond a certain size, i.e., ${\sim}3R(E)$, as more H is added to the interior where its molar volume is much smaller than that for H in atmosphere.
Therefore, it is feasible that the sharp decrease in the planets' occurrence at ${\sim}3R(E)$ (i.e., radius cliff) is related to the chemical process we identified in this study.
An important difference from Kite et al.\  \cite{kite2019Superabundance} is that they assumed the physical mixing between H$_2$ and magma for the radius cliff.
It is, however, important to note that they extrapolated experimental results measured below 3~GPa \cite{hirschmann2012Solubility}, accuracy of which is debatable \cite{schlichting2022Chemical}.
Our experiments conducted directly at the pressure conditions relevant to the larger sub-Neptunes' interiors show that the chemical mixing through the Mg--H bond formation could be an important process to consider for the possible enhancement of the H ingassing at the radius cliff.

If the reaction we identified in this study also occurs at lower pressures, it would be important to consider such chemical interactions in early rocky planets with H-rich primary atmosphere \cite{young2023Earth}.
It is of particular interest how the potential formation of hydride minerals on the surface can impact the habitability of the early rocky planets.

The on-going space telescope (e.g., JWST) and large ground-based telescope missions aim to measure the atmosphere chemistry of exoplanets (including sub-Neptunes) which will be an important step toward answering a fundamental question: are we alone in the universe \citep{decadalsurveyonastronomyandastrophysics2020astro20202021Pathways}? 
Because hydrogen-rich atmosphere should be reactive particularly with refractory materials (silicates and oxides) underneath, the atmosphere-interior interaction is key to understand such astrophysical measurements. 
However, models so far considered mostly physical mixing of H$_2$ (or H$^0$) in silicate magma \citep{hirschmann2012Solubility,olson2018Hydrogen}. 
Latest experiments begin to reveal other important interactions between hydrogen-rich atmosphere and magma: H$_2$O formation through the Fe$^{2+}$ reduction \citep{horn2023Reaction}, and H$^{0}$ alloying with Fe metal \citep{piet2023Superstoichiometric}. 
Here we demonstrated that anionic hydrogen, H$^{-}$ (or hydride), can exist in the setting expected for sub-Neptune exoplanets.
Furthermore, our experiments reveal that hydride can form together with water, both of which can partition to the atmosphere and the interior of sub-Neptunes (Fig.\,\ref{5_Density}).
Therefore, it is important to consider these reactions in the interpretation of exoplanet atmosphere spectra. 
In fact, a recent astrophysical study detected CrH in the H-rich atmosphere of hot Jupiter WASP-31b together with H$_2$O \citep{braam2021evidence}, demonstrating the importance of understanding the type of chemical interaction we report here even for the less abundant heavy elements. 
Hydrides have been important materials to study in condensed matter physics for hydrogen storage \citep{sakintuna2007Metal} and high-temperature superconductivity \citep{duan2017Structure}. 
Our study demonstrates that such studies of hydrides can also impact our understanding on the evolution and dynamics of sub-Neptune exoplanets, which are common in our galaxy \citep{bean2021Nature}.

\showmatmethods{} 

\acknow{
We thank two anonymous reviewers for their constructive criticism and helpful comments, which improved our paper.
This work was supported by National Science Foundation (NSF) grants AST2108129 (T.K. and S.-H.S.) and AST2005567 (X.W. and S.-H.S). 
This work was performed at GSECARS (The University of Chicago, Sector 13), APS, Argonne National Laboratory (ANL). 
GSECARS is supported by the NSF - Earth Sciences (EAR-1634415). 
This research used resources of the APS, a U.S. Department of Energy (DOE) Office of Science User Facility operated for the DOE Office of Science by ANL under Contract No. DE-AC02-06CH11357.
Use of the GSECARS Raman Lab System was supported by the NSF MRI Proposal (EAR-1531583).
We also acknowledge the use of facilities at the Eyring Materials Center at ASU supported in part by NNCI-ECCS-1542160\@.
}

\showacknow{} 

\bibliography{230526_ver,add-shim}

\end{document}